\documentclass[12]{article}

\usepackage{amssymb,latexsym,amsmath}
\usepackage{amsmath}     % Standard packages
\usepackage{dsfont}
\usepackage{upgreek}
\usepackage{graphics}
\usepackage{graphicx}
\usepackage{lscape}
\usepackage{multirow}
\usepackage{amssymb}
\usepackage{dsfont}
\usepackage{hyperref}
\usepackage{authblk}
\usepackage[switch,columnwise]{lineno}
\usepackage{graphicx}
\usepackage{caption}
\usepackage{subcaption}
\usepackage{booktabs}

\usepackage[dvipsnames]{xcolor}
\usepackage{multirow}

\usepackage[margin=1.5in]{geometry}

\begin{document}
\title{Weighted geometric distribution with a new characterisation of geometric distribution}
\author{\textbf{Deepesh Bhati$^{\dagger*}$ and Savitri Joshi}
\footnote{Department of Statistics, Central University of Rajasthan, Kishangarh-305817,\\ \indent $^\dagger$deepesh.bhati@curaj.ac.in(corresponding author)}
}

\maketitle
\begin{abstract}
\noindent In this paper, we introduce a new generalization of geometric distribution which can also viewed as discrete analogue of weighted exponential distribution introduced by Gupta and Kundu(2009). We derive some distributional properties like moments, generating functions, hazard function and infinite divisibility followed by three methods of estimation of the parameters. A new characterisation of Geometric distribution have also been presented using the proposed distribution. Finally, we examine the model with real data sets.
\end{abstract}

\section{Introduction}
Recently, count data modelling becomes very popular in many areas like Insurance, Ecology, Environmental Science, Health etc., because of larger variance is than mean. To model such situations standard models like Poisson , negative binomial, geometric distribution are not enough, which makes researchers to generalize, extend or modify these standard models preserving the fundamental properties like uni modality, log concavity, infinite divisibility, over dispersion(under dispersion) of standard distributions. Some of such generalization you can find in Jain and Consul(1971), Philippou and Georghiou(1983), Tripahti et al.(1987). However researchers still continuing to propose new generalization of these standard distributions. Chakraborty (2015) gave a comprehensive survey on method and construction of generating discrete analogue of continuous distribution. \\
Because of the applications and elegant and mathematical tractable distributional form geometric distribution attracts many research for count data modeling. In last 6 years, many generalization of geometric distribution have been appear in the literature using mainly three methodology, (i) Discretizing the continuous distribution (ii) Mixed Poisson(mixing poisson parameter with continuous distribution) Methodology and (iii) taking discrete Analogue of continuous distribution over real line.(see definition () in section). Connection of many generalized geometric distributions obtained by utilizing these three methods  have been briefly presented in Figure 1,2 and 3.

\begin{figure}
\begin{center}
\includegraphics[scale=.5]{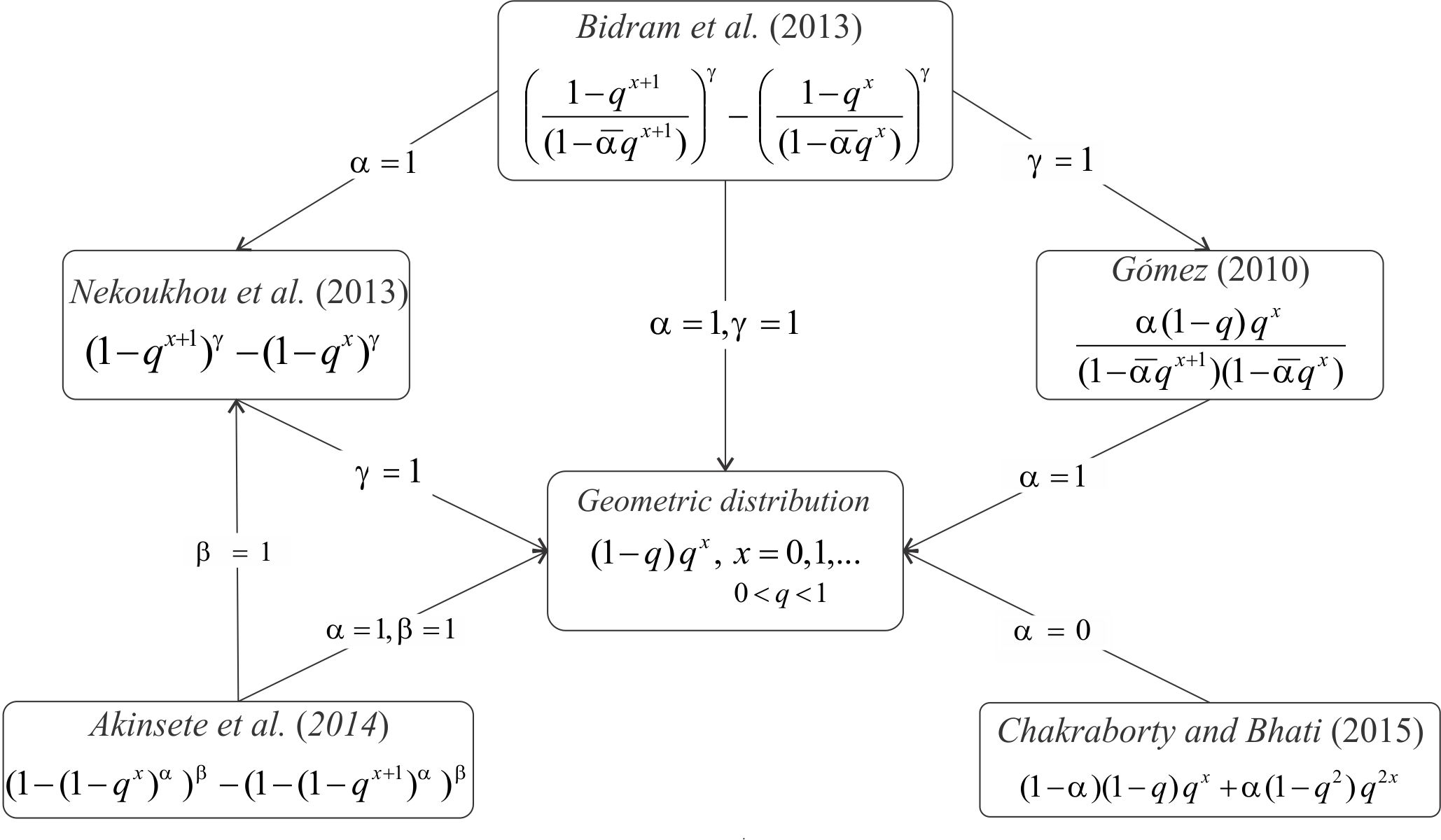}
\caption{Connection of different generalized geometric distribution, obtained by discretization method, with standard geometric distribution.}
\end{center}
\end{figure}

\begin{figure}
\begin{center}
\includegraphics[scale=.5]{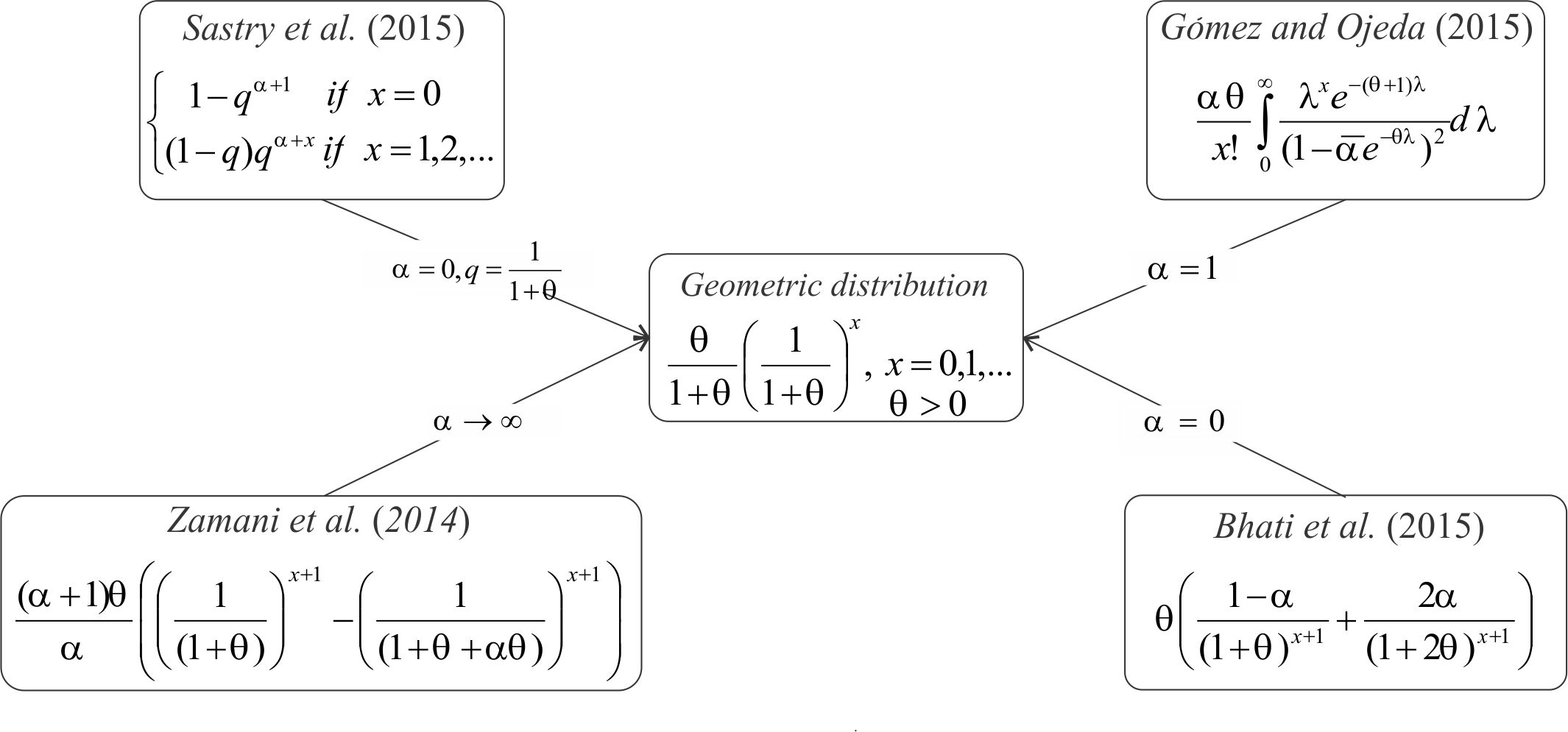}
\caption{Connection of different generalized geometric distribution, obtained by Mixed-Poisson method, with standard geometric distribution.}
\end{center}
\end{figure}

\begin{figure}
\begin{center}
\includegraphics[scale=.5]{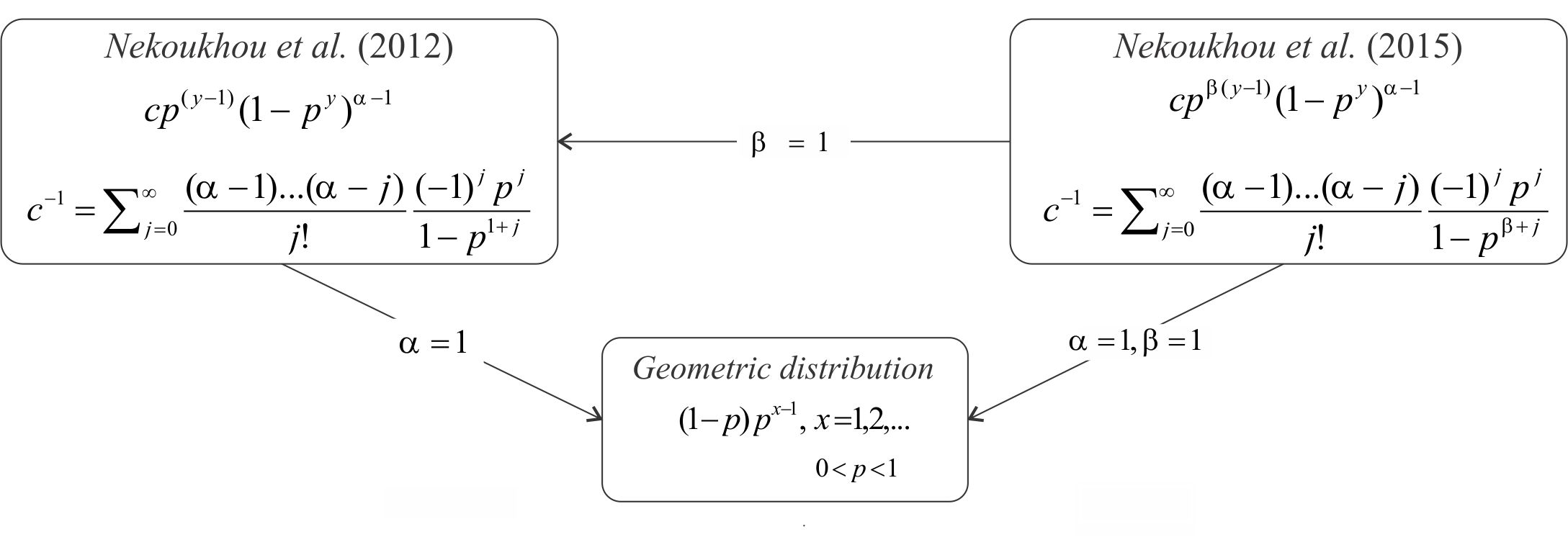}
\caption{Connection of different generalized geometric distribution, obtained by discrete analogue method, with standard geometric distribution.}
\end{center}
\end{figure}

In this article we propose another generalization of \textit{``geometric distribution''} and possessing many properties like, unimodality, log concavity, infinite divisibility, closed form of moments estimator and estimator obtained using proportion of zeros and mean. Moreover the proposed distribution can also be viewed as discrete analogue of weighted exponential distribution by Gupta and Kundu(2009) and Weighted version of geometric distribution (see Patil and Rao(1977, 1978), Rao(1985)), in fact we proposed our distribution using the later. The weight function proposed in this paper is $w(x)=1-q^{\alpha(x+1)}$, where $\alpha$ is the shape parameter the parameter $q$ comes from the definition of geometric distribution. The theoretical and practical aspect of this new family is discussed. Interestingly, from the proposed distribution, some new characterizations for Geometric Distribution are shown. The paper is structured as follows, after definitions and basic distributional properties discussed in section 2, we present various results associated with proposed model in section 3.  Then in Section 4 we present some characterisations of geometric distribution from the proposed model. In section 5 we consider statistical issues of parameter estimation. In the subsequent section it is shown that generated models give good fit for real data set.

\section{Definition and Distributional Properties}
Let $X$ be a geometric rv defined as $p_X(x)=(1-q)q^x,$ for $x=0,1,2,3,\cdots$, then for non-negative weight function $w(x)=1-q^{\alpha(x+1)}$, let $Y$ represents the weighted geometric rv having PMF
\begin{equation} \label{e1}
p_Y(y)=\frac{w(y)p_X(y)}{E(w(X))}=\frac{(1-q^{\alpha(y+1)})(1-q)q^y}{E(1-q^{\alpha(X+1)})}
\end{equation}
where
\[
E(W(X))=E(1-q^{\alpha(X+1)})=\sum\limits_{x=0}^{\infty} (1-q^{\alpha(x+1)})(1-q)q^x=\frac{1-q^\alpha}{1-q^{\alpha+1}}
\]
Thus, from (\ref{e1}) we have
\begin{equation} \label{e2}
p_Y(y)= \frac{(1-q)(1-q^{\alpha+1})}{(1-q^{\alpha})}q^{y}(1-q^{\alpha(y+1)})  \quad;\quad y=0,1,\cdots
\end{equation}
Henceforth, we call rv defined in (\ref{e2}) as \textit{Weighted Geometric Distribution} and denote it as $Y \sim \mathcal{WG}(\alpha,q)$, where $0<q<1$ and $\alpha>0$. It can be observed straightly that as $\alpha \rightarrow \infty $, $q^{\alpha} \rightarrow 0$, hence the proposed PMF(\ref{e2}) reduces to geometric distribution with PMF
\begin{equation*}
p_y=(1-q)q^y, \qquad y \in \mathcal{N}_0
\end{equation*}
Further the ratio of the two probabilities can be written as
\begin{equation} \label{e3}
\frac{p_{y+1}}{p_y}=\frac{q \left(1-q^{\alpha(y+2)}\right)}{\left(1-q^{\alpha(y+1)}\right)}, \qquad  y=0,1, \cdots
\end{equation} 
with $p_0=(1-q)(1-q^{\alpha+1})$.

As the parameters $0<q<1$ and $\alpha>0$, we can observe that, $1-q^{\alpha(y+1)}<1-q^{\alpha(y+2)}<\cdots$, increases with $y$ so that the ratio of probabilities in (\ref{e3}) decreases with increase in $y$, and hence the distribution is unimodal. Moreover, because $p^2_y>p_{y+1}p_{y-1}$ for all $y$, the distribution is log-concave and strongly unimodal (see Keilson and Gerber 1971). Therefore,  $\mathcal{WG}(\alpha,q)$ has all moments. For different values of parameters $\alpha$ and $q$, the PMF of $\mathcal{WG}(\alpha,q)$ is presented in Figure 1.\\
As it has been noted that, $\mathcal{WG}(\alpha,q)$ is log-concave and unimodal, it implies that there exist a variable $y^*$ (mode) that satisfies the following equation 
\begin{eqnarray*} \label{}
\begin{cases} p_{y+1} \ge p_y \qquad & \forall \quad y\le y^*, \\
p_{y+1} \le p_y \qquad & \forall \quad y\ge y^*,
\end{cases}
\end{eqnarray*}
hence under certain constraints on parameters we obtained the mode of $\mathcal{WG}(\alpha,q)$ as follows, for $0<q<1/2$ and $\alpha>0$ the mode of $\mathcal{WG}(\alpha,q)$  is at origin, whereas for $1/2<q<1$ and $\alpha> \log_q\left(\frac{1-q}{q}\right)$ the mode is at $\left[\log_q\left(\frac{1-q}{1-q^{\alpha+1}} \right)^{1/\alpha}\right]$. If $\left[\log_q\left(\frac{1-q}{1-q^{\alpha+1}} \right)^{1/\alpha}\right]$ is an integer, then there are joint modes at $\left[\log_q\left(\frac{1-q}{1-q^{\alpha+1}} \right)^{1/\alpha}\right]$ and $1+\left[\log_q\left(\frac{1-q}{1-q^{\alpha+1}} \right)^{1/\alpha}\right]$. Here $[.]$ denotes the integral part. \\

\begin{figure}
\begin{center}
\includegraphics[scale=.50]{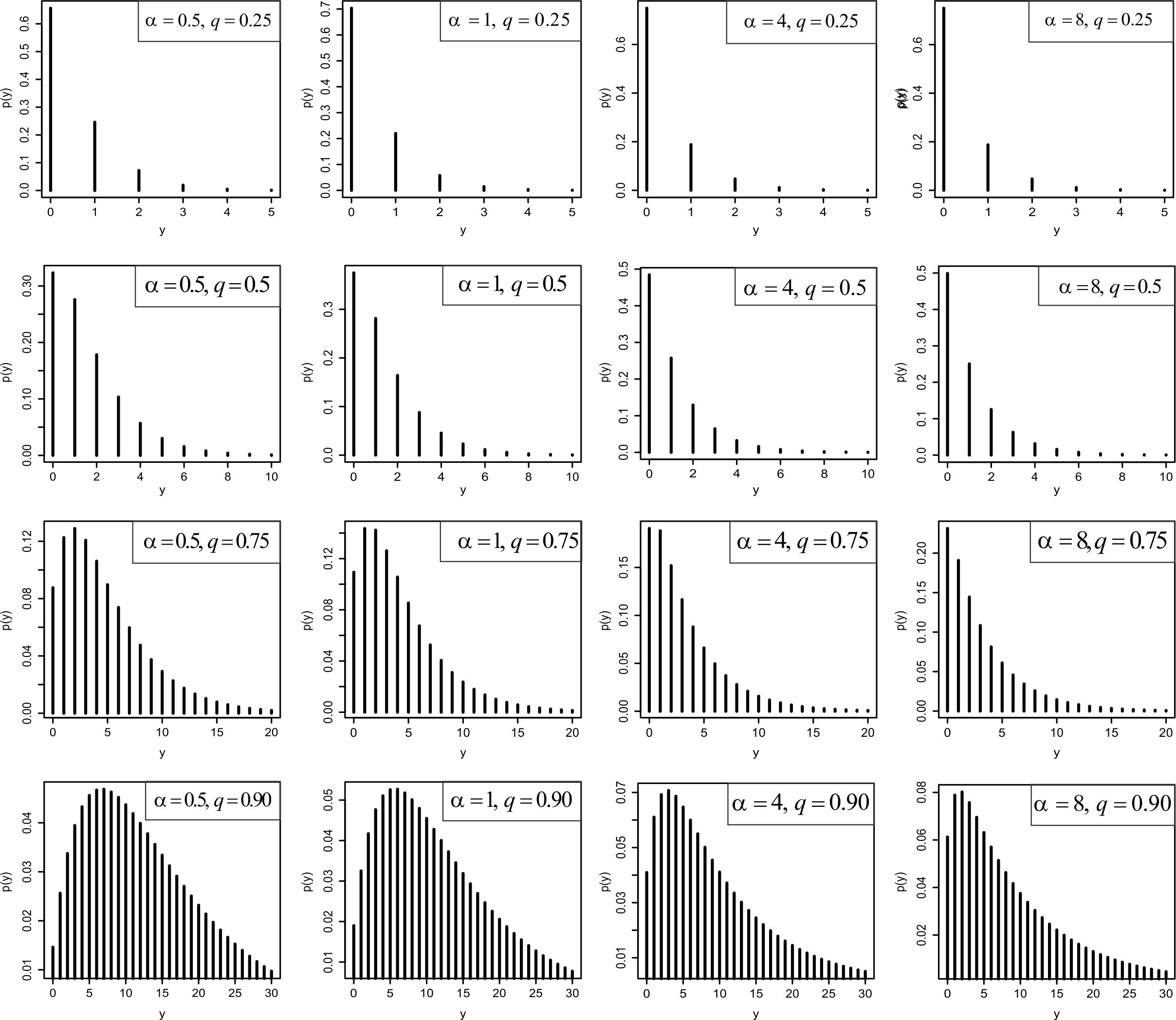}
\caption{PMF plot of $\mathcal{WG}$ distribution for different values of parameters $\alpha$ and $q$.}
\end{center}
\end{figure}

\noindent Moreover, by using series expansion(see, e.g., Graham et al., 1989, formula (7.46), p. 337) 
\begin{equation*}
\sum\limits_{y=0}^{\infty}y^nq^y= \sum\limits_{y=1}^{\infty}S(n,y)\frac{y!q^y}{(1-q)^{1+y}} 
\end{equation*}
\noindent where, 
\begin{equation*}
S(n,y)=\frac{1}{y!}\sum\limits_{j=0}^{y-1}(-1)^j\binom{y}{j}(y-j)^n
\end{equation*}
is the Stirling number of the second kind, the $n^{th}$ raw moment of $\mathcal{WG}(q,\alpha)$ is as follows
\begin{align*}
\mathbb{E}(Y^n)=&\frac{(1-q)(1-q^{\alpha+1})}{1-q^\alpha}\sum\limits_{y=0}^{\infty} y^n q^y\left(1-q^{\alpha(y+1)}\right) \\
=&\frac{(1-q)(1-q^{\alpha+1})}{1-q^\alpha}\left(\sum\limits_{y=0}^{\infty} y^n q^y -q^{\alpha}\sum\limits_{y=0}^{\infty} y^n q^{(\alpha+1)y}\right)\\
=&\frac{(1-q)(1-q^{\alpha+1})}{1-q^\alpha} \left(\sum\limits_{y=1}^{\infty}S(n,y)\frac{y!q^y}{(1-q)^{1+y}}-q^\alpha \sum\limits_{y=1}^{\infty}S(n,y)\frac{y!q^{(\alpha+1)y}}{(1-q^{\alpha+1})^{1+y}} \right) 
\end{align*}
Thus the mean and variance of $\mathcal{WG}(q,\alpha)$ are as follows. 
\begin{equation}
\begin{aligned} \label{e9}
\mathbb{E}(Y)=&-2+\frac{1}{1-q}+\frac{1}{1-q^{1+\alpha}} \\ 
\mathbb{V}(Y)=& q\left(\frac{1}{(1-q)^{2}}+\frac{q^{\alpha}}{(1-q^{1+\alpha})^{2}} \right) 
\end{aligned}
\end{equation}
\noindent Since 
\begin{equation*}
\begin{aligned}
\frac{\partial}{\partial q}\mathbb{E}(Y)&=\frac{1}{(1-q)^2}+\frac{q^{\alpha } (1+\alpha )}{\left(1-q^{1+\alpha }\right)^2}>0, \\
\frac{\partial}{\partial \alpha}\mathbb{E}(Y)&=\frac{q^{1+\alpha}\log(q)}{\left(1-q^{1+\alpha}\right)^2} <0
\end{aligned}
\end{equation*}
\noindent Thus mean of $\mathcal{WG}(\alpha,q)$ distribution decreases as $\alpha$ increases and increases as $q$ increases. \\

\noindent The moment generating function(MGF) of a rv $Y \sim \mathcal{WG}(\alpha,q)$ can easily be computed from definition $\mathbb{M}_{Y}(t)=\mathbb{E}(e^{Yt})$ and is given by following expression.
\begin{equation} \label{e8}
\mathbb{M}_{Y}(t)=\frac{(1-q)(1-q^{1+\alpha})}{(1-e^{t}q)(1-e^{t}q^{1+\alpha})}
\end{equation}

\noindent The survival and hazard rate function of  $\mathcal{WG}(\alpha,q)$ is given by the following expressions respectively:
\begin{equation} \label{e4}
\bar{F}_Y(y;\alpha,q)=\frac{q^y \left(1-q^{(1+y) \alpha }-q^{1+\alpha }+q^{1+\alpha +y \alpha }\right)}{1-q^{\alpha }}
\end{equation}
and 
\begin{equation} \label{e5}
h(y;\alpha,q)=\frac{(1-q)\left(1-q^{\alpha +1}\right)\left(1-q^{\alpha(y+1)}\right)}{\left(1-q^{(1+y) \alpha }-q^{1+\alpha }+q^{1+\alpha +y \alpha }\right)}
\end{equation}
It can be seen that the hazard function is an increasing function of $y$ and approaches  $(1-q)$ as $y \rightarrow \infty$. Moreover, hazard function increases with faster rate for lower value of $q$ as well as for higher value of $\alpha.$ \\
\\
The reversed hazard rate function (RHR) and the Second failure rate function(SRF)are given by the following equations respectively:
\begin{equation} \label{e6}
h^{*}(y)= \frac{P(Y=y)}{F_{Y}(y)} =\frac{(q-1)q^y(-1+q^{\alpha+1})(1-q^{\alpha(y+1)})}{-1+q^{\alpha}+q^y(1-q^{\alpha+1}+(q-1)q^{\alpha+y{(\alpha+1)}}}
\end{equation} \\
and
\begin{align}
\nonumber r^{*}(y)=&\log\left\lbrace\frac{S_{Y}(y)}{S_{Y}(y+1)}\right\rbrace \\
=&\log\left\lbrace\frac{-1+q^{\alpha+1}+(1-q)q^{\alpha(1+y)}}{-1+q^{\alpha+1}+(1-q)q^{\alpha(2+y)}}\right\rbrace
\end{align}

\section{Some Results}
\noindent \textbf{Theorem 1:} The negative binomial distribution $\mathcal{NB}(2,q)$ will be a limiting case of $\mathcal{WG}(\alpha,q)$ distribution as $\alpha \rightarrow 0$.\\
\textit{Proof:} The proof is straight forward after trivial steps. \\

\noindent In last few decades, many methods were proposed in the literature to define discrete distributions which are discrete analogue of continuous rv(comprehensive survey of these methods can be seen in Chakraborty(2015)). Thus in order to explore the connection between continuous distribution with $WGD(q,\alpha)$, we make use of following method define as follows\\

\noindent \textbf{Definition 1:} A rv $Y$ is said to be discrete analogue of continuous rv $X$ if its PMF is defined as 
\[p(y)=f_X(y)/\sum\limits_{r=0}^{\infty}f(r) \qquad  \text{for} \quad y=0,1,\cdots \]

\noindent \textbf{Theorem 2:} If $X \sim \mathcal{WED}(\alpha,\lambda)$, then the rv $Y$ using definition(1), follows $\mathcal{WG}(\alpha,e^{-\lambda})$ distribution. \\
\textit{Proof:} Let $X \sim \mathcal{WED}(\alpha,\lambda)$ with pdf
\[
f_X(x)=\frac{\alpha}{\alpha+1}\lambda e^{-\lambda x}(1-e^{\alpha \lambda x}) \qquad \text{for} \quad x>0
\]
then the discrete analogue of $X$ with PMF
\[
p(y)=\frac{f_X(y)}{\sum\limits_{r=0}^{\infty}f_X(r)}= \frac{e^{-\lambda y}(1-e^{\alpha \lambda y})}{\sum\limits_{r=0}^{\infty}e^{-\lambda r}(1-e^{\alpha \lambda r})} =\frac{(1-e^{-\lambda})(1-e^{-\lambda(\alpha+1)})}{(1-e^{-\lambda\alpha})} e^{-\lambda y}(1-e^{\alpha \lambda y})
\]
Hence by assuming $e^{-\lambda}=q$, we get the result. \hfill  $\Box$\\

\noindent \textbf{Stochastic representations:} In the following theorem, stochastic representations of $\mathcal{WG}(q,\alpha)$ distribution are shown, which can be useful for generating random number and characterization results we will be using later \\ 

\noindent \textbf{Theorem 3:} Let $Y_i, i=1,2$ be independent and identically distributed(iid) rvs with PMF, $p_{Y}(y)$ and distribution function (DF) $F_{Y}(y)$. For any positive integer $\alpha$, let $Y$ denotes the conditional rv $Y_1$ given $\left(\alpha(Y_{1}+1)\ge Y_{2}+1\right)$. Then
\begin{enumerate}
\item[(a)] the PMF of $Y$ is 
\begin{equation} \label{e10}
p_{Y}(y)=\frac{1}{P\left(Y_{2}+1\le\alpha(Y_{1}+1)\right)}p_{Y_1}(y)F_{Y_2}\left(\alpha(y+1)-1\right); \quad\quad y=0,1,\cdots
\end{equation}
\item[(b)] $Y\sim \mathcal{WG}(\alpha,q)$ distribution if $Y_{i}, i=1,2$ are iid geometric rv.
\end{enumerate}

\noindent \textit{Proof:} $(a)$ Let $Y_{1}$ and $Y_{2}$ are two i.i.d. discrete rv 
\begin{align} \nonumber
\nonumber p_{X}(y)=& P(Y_{1}=y|Y_{2}+1\leq \alpha(Y_{1}+1)) \\
\nonumber =&\frac{P(Y_{1}=y \bigcap Y_{2}+1\leq\alpha(Y_{1}+1))}{P(Y_{2}+1\leq \alpha(y_{1}+1))} \\
\nonumber =&\frac{P(Y_{1}=y)P(Y_{2}+1\leq \alpha(y+1))} {P(Y_{2}+1\leq \alpha(Y_{1}+1))} \\
=&\frac{1}{P(Y_{2}+1\le\alpha(Y_{1}+1))}p_{Y_1}(x)F_{Y_2}(\alpha(y+1)-1) \label{e11}
\end{align}
 $(b)$ Considering $Y_i \sim Geo(q)$ for $i=1,2$, and using theorem $3(a)$, we get
\begin{equation} \label{e12}
p_{Y}(y)=\frac{(1-q)q^{y}(1-q^{\alpha(y+1)})}{P(Y_{2}+1\leq \alpha(Y_{1}+1))}
\end{equation}
further, 
\begin{equation*}
\begin{aligned}
P(Y_{2}+1\leq \alpha(Y_{1}+1))=&\sum_{x=0}^{\infty} \sum_{y=0}^{\alpha(x+1)-1}(1-q)q^{x}(1-q)q^{y} \\
=&(1-q)^{2}\left(\sum_{x=0}^{\infty}q^{x}
\left(\sum_{y=0}^{\alpha(x+1)-1}q^{y}\right)\right) \\
=&(1-q)\left(\sum_{x=0}^{\infty}q^{x}(1-q^{\alpha(x+1)})\right) \\
=& (1-q)\left(\sum_{x=0}^{\infty}q^{x}-\sum_{x=0}^{\infty}
q^{x(\alpha+1)+\alpha}\right) \\
=& (1-q)\left(\frac{1}{1-q}-\frac{q^{\alpha}}{1-q^{\alpha+1}}\right) \\
P(Y_{2}+1\leq \alpha(Y_{1}+1))=& \left( \frac{1-q^{\alpha}}{1-q^{\alpha+1}}\right),
\end{aligned}
\end{equation*}
substituting the above result in \ref{e12}, we get the desired result. \hfill $\Box$\\

\noindent \textbf{Theorem 4:} Suppose $X$ and $Z$ be two dependent rvs with joint PMF given as
\begin{equation} \label{e13}
p_{X,Z}(x,z)=
\begin{cases}
(1-q)(1-q^{z})q^{z(x+1)} \,\,\, & \text{for} \,\, x=0,1,2,\cdots, \\
& \quad \,\,\, z=1,\cdots \\
0 & \text{otherwise}
\end{cases}
\end{equation}
Then rv $Y$ defined as $Z-1|X<\alpha$ follows $\mathcal{WG}(\alpha,q)$ distribution. \\
\noindent \textit{Proof:} Let $X$ and $Z$ are two discrete random variables, where $X$ depends on $Z$, with the joint PMF as given below
\begin{equation*}
p_{X,Z}(x,z)=(1-q)(1-q^{z})q^{z(x+1)} \quad;\quad x=0,1,2, \text{and} \quad z=1,2,\cdots
\end{equation*}
Then the marginal probability mass function of $X$  is given as
\begin{align} \label{e14}
\nonumber P_{X}(x)=&\sum_{z=1}^{\infty}
(1-q)(1-q^{z}) q^{z(x+1)} \\
=&\frac{1-q}{1-q^{x+1}}-\frac{1-q}{1-q^{x+2}}; \quad x=0,1,\cdots
\end{align}
Further,
\begin{eqnarray} \label{e15}
\nonumber P(Y=y)&=& P(Z=y+1 |X <\alpha) \\
\nonumber &=&\frac{P(Z=y+1 \cap X <\alpha)}{P(X <\alpha))}\\
\nonumber &=&\frac{\sum_{x=0}^{\alpha-1}P(Z=y+1\cap X<\alpha)}{P(X<\alpha)} \\
\nonumber &=&\frac{\sum_{x=0}^{\alpha-1}(1-q)(1-q^{y+1})q^{(y+1)(x+1)}}{P(X<\alpha)}  \\
&=& \frac{(1-q)(1-q^{\alpha(y+1)})q^{y+1}}{P(X<\alpha)} 
\end{eqnarray}
\noindent using (\ref{e14}),
\[
P(X<\alpha)=\sum_{x=0}^{\alpha-1}P(Y=y)=\sum_{y=0}^{\alpha-1}\left(\frac{1-q}{1-q^{1+x}}-\frac{1-q}{1-q^{2+x}}\right)=	\frac{q(1-q^{\alpha})}{(1-q^{\alpha+1})}
\]
hence, by substituting in (\ref{e15}), we get
\begin{flalign*}
   && P(Y=y)= \frac{(1-q)(1-q^{\alpha+1})q^{y}(1-q^{\alpha
   (y+1)})}{(1-q^{\alpha})} && \Box
\end{flalign*}

Analogue to weighted exponential distribution, we can interpreted above as the discrete hidden truncation model. Suppose $Z$ and $X$ are two correlated discrete rvs with the joint PDF (\ref{e13}) and we can not observe $X$, but we can observe $Z-1$ if $X < \alpha$, then the observed sample can be regarded as drawn from a distribution with the PDF given in $(\ref{e2})$. \\

\noindent \textbf{Theorem 5:} If $ U $ and $ V $ are independent geometric rvs distributed with parameter $q$ and $q^{\alpha+1}$ respectively, then $U+V \sim \mathcal{WG}(\alpha,q)$. \\
\textit{Proof:} Since $U$ and $V$ are independent geometric rvs with parameters $q$ and $q^{\alpha+1}$.
Therefore
\begin{align} \nonumber 
\mathbb{M}_{U+V}(t)&=\mathbb{M}_{U}(t)\mathbb{M}_{V}(t)\\
&=\frac{(1-q)(1-q^{1+\alpha})}{(1-e^{t}q)(1-e^{t}q^{1+\alpha})} \label{e16}
\end{align}
which is same as the MGF of $\mathcal{WG}(\alpha,q)$ given in equation $(\ref{e8})$.
Hence by uniqueness theorem of moment generating functions we conclude that $U+V \sim \mathcal{WG}(\alpha,q)$. \\

\noindent \textbf{Corollary 1:} If $U_i \overset{i.i.d.}{\sim} Geo(q^{1/n})$ and $V_i \overset{i.i.d.}{\sim} \textit{Geo}(q^{\frac{\alpha+1}{n}})$ be two independent samples, then 
\begin{equation*}
U_{(1)}+V_{(1)} \sim \mathcal{WG}(q,\alpha)
\end{equation*}
\noindent where $U_{(1)}=\min_{i\le 1 \le n}U_i$ and $V_{(1)}=\min_{i\le 1 \le n}V_i$. \hfill $\Box$\\
\\
\noindent The result of Theorem 5, can further be extended to obtained new generalized negative binomial distribution, and for its construction, let $X_1$ and $X_2$ follows independent Negative Binomial with parameters $(r,q)$ and $(r,q^{\alpha+1})$ respectively, then the pmf of rv $Y=X_1+X_2$ obtained using convolution definition as follows
\begin{align*}
P(Y=y)&=\sum\limits_{x=0}^{y}p_{X_1}(y-x)p_{X_2}(x),   \text{for} \quad  y=0,1,2,\cdots \\
&=\sum\limits_{x=0}^{y}\binom{r+x-1}{x}\binom{r+y-x-1}{z-x}(1-q)^r q^{y-x}(1-q^{\alpha+1})^r(q^{\alpha+1})^{x}\\
&=(1-q)^r(1-q^{\alpha+1})^r q^{y} \sum\limits_{x=0}^{y}\binom{r+x-1}{x}\binom{r+y-x-1}{y-x} q^{\alpha x}\\
P(Y=y)&=(1-q)^r(1-q^{\alpha+1})^r q^{y} \binom{r+y-1}{y} \,_2F_1\left(r,-y;-r-y+1;q^{\alpha}\right)
\end{align*}
We can call the above PMF as \textbf{weighted negative binomial distribution} with parameters $(r,\alpha,q).$

\subsection {Infinite Divisibility} 
Since the geometric distribution is infinitely divisible, hence it can be seen, in light of theorem 5, that $Y \sim \mathcal{WG}(\alpha,q)$ distribution is also infinitely divisible. Moreover, by factorization property of geometric law(feller,1957) 
\[
Y \stackrel{D}{=}Y_{n1}+Y_{n2}+\cdots+Y_{nn} \qquad \qquad n=1,2,\cdots
\]
where $Y_{ij}=U_{i1}+U_{i2}$ where $U_{i1}$ and $U_{i2}$ are independent negative binomial random variate with parameters $(1/n,q)$ and $(1/n,q^{\alpha+1})$ respectively.
\noindent A canonical representation of $\mathcal{WG}$ characteristic funciton(c.f.) follows from L\'evy-Khintchine theorem stated as 
A complex values function $\varphi$ defined on $\mathbb{R}$ is an infinite divisible c.f. iff $\log \varphi$ admits the representation
\begin{equation} \label{ecf}
\log \varphi(t)=i \gamma t+\int\limits_{-\infty}^{\infty}\left(e^{i t x}-1-\frac{itx}{1+x^2}\right)\frac{1+x^2}{x^2}dG(x)
\end{equation}
where $\gamma \in \mathbb{R}$, $G$ is bounded, non-decreasing, right continuous function on $\mathbb{R}$ such that $G(-\infty)=0$ and $G(\infty)<\infty$.\\
\noindent According to (\ref{e16}),c.f. of $\mathcal{WGD}$ can be written as 
\begin{equation*}
\varphi(t)=\mathbb{M}_Y(i t)= \frac{(1-q)(1-q^{1+\alpha})}{(1-e^{i t}q)(1-e^{it}q^{1+\alpha})}
\end{equation*} 
which after factorisation represent in the form
\begin{equation} \label{cf}
\log \varphi(t)= \sum\limits_{k=1}^{\infty}(e^{i t k}-1)\frac{q^k+q^{(\alpha+1)k}}{k}.
\end{equation}
Taking $\gamma=\sum\limits_{k=1}^{\infty}\frac{q^k+q^{(\alpha+1)k}}{1+k^2}$ and $G(.)$ to be non decreasing function with jumps of magnitude $\frac{k}{k^2+1}\left(q^k+q^{(\alpha+1)k}\right)$ at $k=0,1,2,\cdots$, we see that $\log \varphi(t)$ has a unique representation as given in (\ref{ecf}). Hence we conclude that $\mathcal{WGD}(q,\alpha)$ is infinitely divisible. 

\section{Characterisation of Geometric Distribution using $\mathcal{WG}(\alpha,q)$ Distribution}

Following three theorems gives an alternate way of characterisation of geometric distribution which can also be viewed as simultaneous characterisation of  $\mathcal{WG}(\alpha,q)$ and geometric distribution.\\

\noindent \textbf{Theorem 6} Suppose $Y_1$ and $Y_2$ are i.i.d. discrete r.v.s with support $N_0$ and for $\alpha \ge 1$, suppose the
conditional r.v. $Y = Y_1|((Y_1+1) \ge \alpha(Y_2 +1))$ follows $\mathcal{WG}(\alpha,q)$ given at (\ref{e2}). Then the distribution of $Y_1$ is \textit{Geo}($q$).\\
\noindent \textit{Proof:} By equation (\ref{e10}) we have for all $\alpha \ge 1$ and $y \in N_0$,
\[
f(y) F(\alpha(y+1)-1)=C(\alpha) \frac{(1-q)(1-q^{\alpha+1})}{(1-q^\alpha)}q^y\left(1-q^{\alpha(y+1)}\right)
\]
where $c(\alpha)=P(Y_2+1 \le \alpha(Y_1+1)).$ In particular for $\alpha =1$ we have for all $y \in N_0$
\begin{equation}
f(y)F(y)=C(1)(1-q^2)q^y(1-q^{y+1}) =C^*q^y(1-q^{y+1})
\end{equation}
Then for $y=0$. we get $f^2(0)=C^*(1-q)$. 

Further for $y=1$, we have $f(1)\left(f(0)+f(1)\right)=C^*q(1-q^2).$ Writing $\theta_1=\frac{f(1)}{f(0)}$, we observe $\theta_1(1+\theta_1)=q(1+q)$. For this quadratic equation in $\theta_1$ the two solutions are $q$ and $-(1 + q)$ of which the second being negative is invalid. Thus $f(1) = qf(0)$.

We claim $f(y) = q^y f(0)$ for all $y \in N_0$. Note that it holds for $y = 0, 1$. Suppose this holds for $y = 1, 2, \cdots n$. We shall prove that it holds for $y = n + 1$. \\
From (19) we have 
\[
f(n+1)F(n+1)=C^*q^{n+1}\left( 1-q^{n+2}\right)
\]
i.e \\
\[
\theta_{n+1}\left( \sum\limits_{y=0}^{n}\theta_y +\theta_{n+1}\right)= \frac{C^*q^{n+1}(1-q^{n+2})}{f^2(0)},
\]
where $\theta_k=\frac{f(k)}{f(0)}$, $k=0,1,\cdots,n+1$. Thus we have by the induction hypothesis,
\begin{equation}
\theta_{n+1}\left( \sum\limits_{y=0}^{n}q_y +\theta_{n+1}\right)=q^{n+1} \left(\sum\limits_{y=0}^{n+1}q^y \right)
\end{equation}

For the quadratic equation (20) in $\theta_{n+1}$, $q^{n+1}$ is a solution and the other solution is  $-\sum\limits_{y=0}^{n+1}q^y$ which is not valid. Thus for all $y \in N_0$, $f(y)=f(0)q^y$. This gives $f(0)=1-q$ and hence the r.v. $Y_1$ follows $Geo(q)$. $\Box$ \\

\noindent \textbf{Theorem 7:} Suppose $Y$ and $Z$ are two rvs such that $Y$ and $Z-1$ are $N_0$ valued with the conditional random
variable $Y$ given $Z = z$ having $Geo(q^Z)$ and the conditional distribution of $Z-1$ given $Y <\alpha $ having $\mathcal{WG}(\alpha,q)$, where $\alpha$ being a positive integer and $0 < q < 1$. Then $Z-1$ follows $Geo(q)$.\\

\noindent \textit{Proof:} By assumption
\[
P(Y=y|Z=z)=(1-q^z)q^{zy}, \qquad \text{for} \quad y=0,1,2,\cdots
\]
and 
\[ P(Z-1=x|Y<\alpha)=A(\alpha,q)q^x(1-q^{\alpha(x+1)})
\] 
where $A(\alpha,q)=\frac{(1-q)(1-q^{\alpha+1})}{(1-q^\alpha)}$. Then for $x\in N_0$
\[
P(Z-1=x|Y<\alpha)=\frac{P(Z=x+1 \cap Y \le \alpha-1)}{P(Y \le \alpha-1)}=A(\alpha,q)q^x(1-q^{\alpha(x+1)})
\]
\begin{eqnarray*}
\frac{P(Z=x+1)P(Y \le \alpha-1|Z=x+1)}{P(Y \le \alpha -1)}&=& A(\alpha,q)q^x(1-q^{\alpha(x+1)}) \\
\frac{P(Z=x+1)(1-q^{\alpha(x+1)})}{P(Y \le \alpha -1)}&=&A(\alpha,q)q^x(1-q^{\alpha(x+1)}) \\
P(Z=x+1)&=&A^*(\alpha,q)q^x 
\end{eqnarray*}
where $A^*(\alpha,q)=A(\alpha,q)P(Y<\alpha)$. Hence $Z-1$ follows $Geo(q)$. \quad \qquad \quad $\Box$\\

\noindent \textbf{Theorem 8} Suppose $Y$ is a rv with support $N_0$ and probability generating function $\mathbb{P}_Y(t)$ and the corresponding weighted distribution with weight function $w(t) = 1-q^{\alpha(t+1)}$ is $\mathcal{WG}(\alpha,q)$, $\alpha \ge 1, 0 < q < 1$. Then $Y$ follows $Geo(q)$. \\
\textit{Proof:} Note that $\mathbb{E}(w(Y))=1-q^\alpha \mathbb{P}(q^\alpha)$. Then the weighted distribution of $Y$ corresponding to the weight function $w(t)$ is given by \\
\begin{equation} \label{e19}
q(y)=\frac{(1-q)(1-q^{\alpha(y+1)})}{1-q^\alpha \mathbb{P}(q^\alpha)} P(Y=y) 
\end{equation}
But by hypothesis
\begin{equation} \label{e20}
q(y)=\frac{(1-q)(1-q^{\alpha+1})}{(1-q^\alpha)}q^y(1-q^{\alpha(y+1)})
\end{equation}
Thus from (\ref{e19}) and (\ref{e20}) we get 
\[
P(Y=y)=C(\alpha,q)(1-q)q^y.
\]
For $Y$ to be a proper r.v. it is necessary that $C(\alpha; q)$ = 1 and hence $Y \sim  Geo(q)$.

\section{Estimation}
In this section we discuss various methods of estimation of parameters $\alpha$ and $q$.
\subsection{Moment Estimators}
The Method of Moments(MM) estimators are obtained by equating the sample mean and sample variance  $M_1=\frac{1}{n}\sum\limits_{i=1}^{n}x_i$ and $M_2=\frac{1}{n}\sum\limits_{i=1}^{n}(x_i-\bar{x})^2$ with the population moments defined in (6) and (7) respectively as
\begin{equation*}
M_1=\frac{\tilde{q}}{1-\tilde{q}}+\frac{\tilde{q}^{{\tilde{\alpha}+1}}}{1-\tilde{q}^{\tilde{\alpha}+1}} , \qquad  M_2=\frac{\tilde{q}}{(1-\tilde{q})^2}+\frac{\tilde{q}^{{\tilde{\alpha}+1}}}{(1-\tilde{q}^{\tilde{\alpha}+1})^2}
\end{equation*}
Hence, by solving the above equations we get moment estimators of $\tilde{q}$ and $\tilde{\alpha}$ in the closed form as
\begin{equation}
\begin{aligned}
\tilde{q}=&\frac{2M_1+M_1^2-M_2+\sqrt{2M_2-M_1^2-2 M_1}}{2+3M_1+M_1^2-M_2} \\
\tilde{\alpha}=& \log_{\tilde{q}}\left( \frac{(1-\tilde{q})M_1-\tilde{q}}{(1-\tilde{q})(M_1+1)-\tilde{q}}\right)-1
\end{aligned}
\end{equation}

\subsection{Estimator based on proportion of zeros and ones(MP)}
If $p_0, p_1$ be the known observed proportion of $0's$ and $1's$ in the sample given as
\begin{equation*}
p_0=(1-\breve{q})(1-\breve{q}^{\breve{\alpha}+1}),\qquad   p_1=(1-\breve{q})(1-\breve{q}^{\breve{\alpha}+1})\breve{q}(1+\breve{q}^{\breve{\alpha}})
\end{equation*}
solving the above equations the closed form solution for $\breve{\alpha}$ and $\breve{q}$ is
\begin{equation}
\begin{aligned}
\breve{q}=&\frac{p_1+\sqrt{4 p_0^2-4 p_0^3-4 p_0 p_1+p_1^2}}{2 p_0} \\
\breve{\alpha}=& \log_{\breve{q}}\left(\frac{p_1}{\breve{q}p_0}-1 \right) 
\end{aligned}
\end{equation}

\subsection{Maximum Likelihood Estimation}
Let $(x_1,x_2,\cdots,x_n)$ be $n$ random observations from $\mathcal{WG}(\alpha,q)$. Then the $\log$-likelihood function$(l=\log L(q,\alpha |x))$ for this sample is given by
\begin{equation}
l= n \log(1-q)+n \log(1-q^{\alpha+1})-n \log(1-q^{\alpha})+\log{q}\sum_{i=1}^{n}x_i +\sum_{i=1}^{n} \log(1-q^{\alpha(x_i +1)})
\end{equation}
Therefore the $\log$-likelihood equations are given as follows.
\begin{equation} \label{e25}
\frac{\partial l}{\partial q}=-\frac{n}{1-q}+\frac{nq^{\alpha-1}\alpha}{1-q^{\alpha}}-\frac{nq^{\alpha}(1+\alpha)}{1-q^{1+\alpha}}+\frac{\sum_{i=1}^{n}x_i}{q}-\sum_{i=1}^{n}\frac{q^{\alpha(1+x_i)-1}\alpha (1+x_i)}{1-q^{\alpha (1+x_i)}}
\end{equation}
\begin{equation} \label{26}
\frac{\partial l}{\partial \alpha}=\frac{n q^{\alpha} \log q}{1-q^{\alpha}}-\frac{n q^{1+\alpha}\log q}{1-q^{1+\alpha}}-\sum_{i=1}^{n}\frac{q^{\alpha(1+x_i)}(1+x_i)\log q}{1-q^{\alpha(1+x_i)}}
\end{equation}
As the above equations could not give closed solution for the parameter. We solve the above equation by numerical methods using the initial value of parameter obtained from either Moment estimator or by Method of Proportion.
\\
The second order partial derivatives are given as follows
\begin{align*}
\frac{\partial^{2} \log L}{\partial q^{2}}&=\left. -\frac{n}{(1-q)^2}+\frac{nq^{\alpha-2}(\alpha-1)\alpha}{1-q^{\alpha}}+\frac{nq^{-2+2\alpha}\alpha^{2}}{(1-q^{\alpha})^2}\right.\\
& \quad \left.-\frac{nq^{\alpha-1}\alpha (1+\alpha)}{1-q^{1+\alpha}}-\frac{nq^{2\alpha}(1+\alpha)^2}{(1-q^{1+\alpha})^2}-\frac{\sum_{i=1}^{n}x_i}{q^2} \right.\\
&\quad \left. +\sum_{i=1}^{n}\left( -\frac{q^{-2+2\alpha(1+x_i)}\alpha^{2}(1+x_i)^2}{(1-q^{\alpha(1+x_i)})}-\frac{q^{-2+\alpha(1+x_i)}\alpha(1+x_i)(-1+\alpha(1+x_i))}{1-q^{\alpha(1+x_i)}}\right) \right. \\
\frac{\partial^{2} \log L}{\partial \alpha^{2}}&= \left. \frac{nq^{2\alpha}\log q^2}{(1-q^{\alpha})^2}+\frac{nq^{\alpha}\log q^{2}}{1-q^{\alpha}}-\frac{nq^{2+2\alpha}\log q^2}{(1-q^{1+\alpha})^2} -\frac{nq^{1+\alpha}\log q^{2}}{1-q^{1+\alpha}}\right.\\
& \quad \left.+ \sum_{i=1}^{n}\left(-\frac{q^{2\alpha(1+x_i)}\log q^{2}(1+x_i)^2}{(1-q^{\alpha(1+x_i)})^2}-\frac{q^{\alpha (1+x_i)}\log q^{2}(1+x_i)^2}{(1-q^{\alpha(1+x_i)})}\right)\right. \\
\frac{\partial^{2}\log L}{\partial q \partial \alpha} &= \left. \frac{n q^{\alpha-1}}{1-q^{\alpha}}-\frac{nq^{\alpha}}{1-q^{1+\alpha}}+\frac{nq^{-1+2\alpha}\alpha \log q}{(1-q^\alpha)^2}+\frac{n q^{-1+\alpha}\alpha \log q}{1-q^{\alpha}}\right. \\
& \quad \left.-\frac{n q^{1+2\alpha}(1+\alpha)\log q}{(1-q^{1+\alpha})^2}-\frac{nq^{\alpha}(1+\alpha)\log q}{1-q^(1+\alpha)} \right. \\
&\quad  +\sum_{i=1}^{n}\left(\frac{q^{-1+\alpha(1+x_i)}(1+x_i)\alpha \log q (1+x_i)^2}{(1-q^{\alpha (1+x_i)})}-\frac{q^{-1+2\alpha(1+x_i)}\alpha \log q (1+x_i)^2}{(1-q^{\alpha(1+x_i)})^2} \right. \\
&\quad  \left.\qquad \quad -\frac{q^{-1+\alpha(1+x_i)}\alpha \log q (1+x_i)^2}{1-q^{\alpha(1+x_i)}}\right)
\end{align*}

\noindent The Fisher's information matrix of $(q,\alpha)$ is 
\begin{equation*}
\mathbf{J_y}= \begin{bmatrix}
 -\mathbb{E}\left(\frac{\partial^2l}{\partial q^2}\right)&-\mathbb{E}\left(\frac{\partial^2l}{\partial q \partial \alpha}\right) \\ 
-\mathbb{E}\left(\frac{\partial^2l}{\partial \alpha \partial q}\right)&-\mathbb{E}\left(\frac{\partial^2l}{\partial \alpha^2}\right) 
\end{bmatrix}
\end{equation*}
which can be approximate and written as  
\begin{equation*} 
\mathbf{J_y} \approx \begin{bmatrix}
J_{11} &J_{12} \\ 
 J_{21}&J_{22} 
\end{bmatrix}=\begin{bmatrix}
 \frac{\partial^2l}{\partial q^2}\big|_{\hat{q},\hat{\alpha}}&\frac{\partial^2l}{\partial q \partial \alpha}\big|_{\hat{q},\hat{\alpha}} \\
\\ 
\frac{\partial^2l}{\partial \alpha \partial q}\big|_{\hat{q},\hat{\alpha}}&\frac{\partial^2l}{\partial \alpha^2}\big|_{\hat{q},\hat{\alpha}} 
\end{bmatrix}
\end{equation*}
where $\hat{q}$ and $\hat{\alpha}$ are the maximum likelihood estimators of $q$ and $\alpha$ respectively. Also, as $n \to \infty$ limiting distribution of $\sqrt{n}\left(q-\hat{q},\alpha-\hat{\alpha}\right)$ distributed as bivariate normal with mean vector 0 and variance-covariance matrix 
\begin{equation}
\mathbf{J}^{-1}=\frac{1}{J_{11}J_{22}-J_{12}J_{21}} \begin{pmatrix}
J_{11} & J_{12}\\ 
J_{21}& J_{22}
\end{pmatrix}
\end{equation}

\section{Data Analysis}
In this section we apply $\mathcal{WG}(q,\alpha)$ distribution to two data sets. The fist data is from Klugman et al.(2012) represents number of claims made by an automobile insurance policyholders. Whereas the second data set is also from (Klugman 2012) gives information about number of hospitalizations per family member per year. Here we apply method of maximum likelihood  to estimate the parameter of our model using maxlik() function in R with initial value taken from method of moment discussed in section(5.1). The variance to mean ratio for dataset 1 and 2 are 1.357 and 1.075  respectively, which give clear sign of over dispersion. Hence, $\mathcal{WG}$ could be one of the possible choices. Further the list of distributions presented in Table (1) which are used for comparative study for data sets. Table (2) and (1) present the details of fitting of $\mathcal{WG}$ and other distributions. Comparative measure like chi-square$\left(\chi^2=\sum_{i}(O_i-E_i)^2/E_i\right)$  test and the corresponding $p-value$ are being used.  \\

\begin{table}[]
\centering
\caption{DetailS of distributions used in data analysis for comparative study.}
\small\addtolength{\tabcolsep}{-4pt}
\begin{tabular}{lr}
\hline
Distribution Name                                  & Distributional form \\ \hline
Negative Binomial $(\mathcal{NB})$                 & \multirow{3}{*}{$\binom{r+y-1}{y}p^r(1-p)^y$}                                             \\
Johnson(2005)                                      &                                                                                           \\
&\\
Poisson Lindley   $(\mathcal{PL})$                 & \multirow{3}{*}{$\frac{\theta^2(y+\theta+2)}{(1+\theta)^3}$}                              \\
Shankaran (1970)                                   &                                                                                           \\
& \\
Generalized Gemometric $(\mathcal{GGD})$           & \multirow{3}{*}{$\frac{\alpha \theta^y(1-\theta)}{\left(1-\bar{\alpha}\theta^y\right)\left(1-\bar{\alpha}\theta^{y+1}\right)}$}                                               \\
G\'omez (2010)                                     &                                                                                            \\
&\\
New Generalized Poisson Lindley $(\mathcal{NGPL})$ & \multirow{3}{*}{$\frac{\theta ^2}{(\theta +\alpha)(1+\theta)^{x+1}}\left(1+\frac{\alpha (x+1)}{(1+\theta )}\right)$} \\
Bhati et al.(2015)                                 & \\
&\\
Poisson -Inverse Gaussian $(\mathcal{PIG})$        & \multirow{3}{*}{$\frac{1}{y!}\sqrt{\frac{2\phi }{\pi}}e^{\phi /\mu } \left(\frac{\phi \mu^2}{2\mu^2+\phi }\right)^{-\frac{1}{4}+\frac{y}{2}}    K_{\frac{1}{2}-y}\left(\sqrt{2\phi +\frac{\phi ^2}{\mu ^2}}\right)$} \\
&\\
Willmot(1987)                                      &\\
&\\
New Discrete Distribution $(\mathcal{ND})$         & \multirow{3}{*}{$\frac{\log(1-\alpha \theta^y)-\log(1-\alpha \theta^{y+1})}{\log(1-\alpha)}$}                                                                                                                        \\
Gomez et al.(2011) & \\
& \\
Discrete Generalized Exponential $(\mathcal{DGE})$ & \multirow{3}{*}{$(1-q^{y+1})^\alpha-(1-q^{y})^{\alpha} $} \\ 
Nekoukhou et al.(2012)                            & \\                           
& \\ \cline{1-2}
\end{tabular}
\end{table}

\begin{table}[]
\centering
  \small\addtolength{\tabcolsep}{-4pt}  
\caption{Number of claims in automobile insurance}
\begin{tabular}{llrrrrr} \hline
 & {Observed} & \multicolumn{5}{c}{Expected frequency} \\  \hline
Count         & frequency        & $\mathcal{PL}$ & $\mathcal{NB}$  & $\mathcal{GGD}$    & $\mathcal{NGPL}$      & $\mathcal{WG}$    \\ \hline
0             & 1563 & 1569.53   & 1564.54       & 1563.67             & 1564.57              & 1564.27          \\
1             & 271  & 256.34    & 264.58         & 266.37               & 264.28              & 265.12           \\
2             & 32   & 41.34     & 39.44         & 38.69               & 39.69               & 39.05            \\
3             & 7    & 6.6       & 5.66          & 5.48                & 5.59                & 5.62             \\
$\ge$4             & 2    & 1.19      & 0.78          & 0.79                & 0.87                & 0.94             \\ \hline
              & 1875     & 1875          & 1875              & 1875                    & 1875                    & 1875                 \\ \hline
\multicolumn{2}{l}{parameter(s)(\textbf{$\Theta$})}       & $\theta$  & $(r, p)$   & ($\alpha$, $\theta$) & ($\alpha$, $\theta$) & ($q$, $\alpha$) \\
\multicolumn{2}{l}{Estimates$(\hat{\Theta})$}           & 5.89      & (1.309, 0.871) & (1.212, 0.141)    & (8.835, 7.874)      & (0.143, 0.874)  \\
\multicolumn{2}{l}{S.E.$(\hat{\Theta})$}                 & 0.32      & (1.081, 0.214) & (0.226, 0.029)     & (2.601, 0.439)      & \textbf{(0.023, 0.707)} \\
\multicolumn{2}{l}{($df$,$\chi^2_{df}$-value)}       & (4, 3.874) & (3, 3.610)     & (3, 3.597)          & (3, 3.489)           & \textbf{(3, 2.935)}       \\
\multicolumn{2}{l}{p-value}             & 0.4231    & 0.4332        & 0.4632              & 0.481                & \textbf{0.569 }\\ \hline
\end{tabular}
\end{table}

\begin{table}[]
\centering
  \small\addtolength{\tabcolsep}{-4pt}
\caption{Number of claims in automobile insurance}
\begin{tabular}{llrrrrr} \hline
 & {Observed} & \multicolumn{5}{c}{Expected frequency} \\  \hline
Count         & frequency        & $\mathcal{ND}$ & $\mathcal{NB}$  & $\mathcal{PIG}$    & $\mathcal{DGE}$      & $\mathcal{WG}$    \\ \hline
0  & 2659 & 2659.02 & 2659.06 & 2658.97 & 2660.62  & 2659.04 \\
1  & 244  & 243.79  & 243.64  & 244.02  & 242.456  & 243.738 \\
2  & 19   & 19.52   & 19.65   & 19.24   & 19.2681  & 19.5579 \\
$\ge$ 3 & 2    & 1.65    & 1.63    & 1.75    & 1.643192 & 1.65801 \\   \hline
     & 2924   & 2924    &2924     & 2924     & 2924      & 2924      \\ \hline
\multicolumn{2}{l}{parameter(s)(\textbf{$\Theta$})}       & $(\alpha,\theta)$  & $(r, p)$   & ($\phi$, $\mu$) & ($\alpha$, $q$) & ($q$, $\alpha$) \\
\multicolumn{2}{l}{Estimates$(\hat{\Theta})$}           & (−0.341,0.079)      & (1.314,0.93) & (0.127,0.098)    & (1.147.079)      & (0.078,0.696) \\
\multicolumn{2}{l}{S.E.$(\hat{\Theta})$}                 & (0.673,0.018)      & (0.652,0.032) & (0.065,0.006)     & (0.256,0.016)  & (0.023,0.984) \\
\multicolumn{2}{l}{($df$,$\chi^2_{df}$-value)}       & (1, 0.08) & (1, 0.09)     & (1, 1.02)          & (1, 0.092)           & (1, 0.08)       \\
\multicolumn{2}{l}{p-value}             & 0.7773    & 0.7641        & 0.3125              & 0.76164                & 0.7773\\ \hline
\end{tabular}
\end{table}

\section{Conclusion}
In this paper, we propose a new generalization of geometric distribution which can also be viewed as a discrete version of weighted exponential distribution. We have derived some distributional properties of the proposed model and also we present three results on characterisation of geometric distribution from the proposed model. Unlike with other generalizations of geometric distribution, closed form expressions for estimation methods viz. method of moment and method of proportion are obtained. Proposed model give better fit for automobile insurance data. Some reliability properties of the proposed model can be looked as further work. Moreover a new extension of negative binomial distribution have also presented in the proposed work, which can further be explored.

\section*{Acknowledgement}
The authors are thankful to Prof. M. Sreehari for bringing our attention to characterisation section and particularly proof of Theorem 6 and to Prof. R. Vasudeva, for his critical reading and discussion.


\begin{thebibliography}{}

\bibitem{} Adrienne W. K.(1997). Characterization of a discrete normal distribution, \textit{Journal of Statistical Planning and Inference}, 63, 223--229.

\bibitem{} Azzalini A.(1985). A class of distributions which includes the normal ones,  \textit{Scandinavian Journal of Statistics}, 12, 171--178.

\bibitem{} Barbiero A.(2013). An alternative discrete skew Laplace distribution, \textit {Statistical Methodology}, 16, 47--67.

\bibitem{} Bidram H., Roozegar R., Nekoukhou V.,(2015) Exponentiated generalized geometric distribution: A new discrete distribution, \textit{Hacettepe Journal of Mathematics and Statistics}, Doi: 10.15672/HJMS.20159013119.

\bibitem{} Chakraborty S.(2015). Generating discrete analogues of continuous probability distributions- A survey of methods and constructions, \textit{Journal of Statistical Distributions and Applications}, 2:6.

\bibitem{} Chakraborty, S., and Bhati D. (2016) Transmuted Geometric Distribution with Applications in Modeling and Regression Analysis of Count Data. (submitted) 

\bibitem{} Finkelstein M.S.(2002). On the reversed hazard rate, \textit {Reliability Engineering and System Safety}, 78, 71--75.

\bibitem{} G\'omez-D\'eniz , E.(2010). Another generalization of the geometric distribution,\textit{ Test}, 19, 399 – 415 .

\bibitem{} G\'omez D. E., and Ojeda E. C.(2011). The Discrete Lindley distribution; properties and applications,\textit{Journal of Statistical Computation and Simulation}, 81(11), 1405 - 1416.

\bibitem{} G\'omez-D\'eniz, E. and   Ojeda E. Calderín (2014): Parameters Estimation for a New Generalized Geometric Distribution, \textit{Communications in Statistics - Simulation and Computation}, DOI: 10.1080/03610918.2013.835410.

\bibitem{} Gupta R.D., and Kundu D.(2009). A new class of weighted exponential distributions, \textit{Statistics}, 43, 621-634.

\bibitem{} Inusah S., Kozubowski T.J., (2006). A discrete analogue of the Laplace distribution, \textit{Journal of Statistical planning and inference}, 136, 1090--1102.

\bibitem{} Jain , G. C. and Consul , P. C. ( 1971 ). A generalized negative binomial distribution. SIAM Journal of Applied Mathematics; 21:501–513.

\bibitem{} Lindley D. V.(1985). Fiducial Distributions and Distributions and Bayes' Theorem, \textit{Journal of Royal Statistical Society}, 20(1), 102--107

\bibitem{} Mak\u{c}utek J (2008) A generalization of the geometric distribution and its application in quantitative linguistics.
\textit{Romanian Rep Phys}, 60(3):501–509.

\bibitem{} Nekoukhou, V., Alamatsaz, M. H. and Bidram, H. (2011). Discrte generalized exponential distribution of a second type. \textit{Statistics}, 

\bibitem{} Patil G. P.(1964). On Certain Compound Poisson  and Compound Binomial  Distributions, \textit{Sankhy\={a} A}, 27, 293--294.

\bibitem{} Patil, G.P.(1981). Studies in statistical ecology involving weighted distributions, in Statistics Applications and New Directions: Proceedings of ISI Golden Jubilee International Conference, J.K. Ghosh \& J. Roy, eds, \textit{Statistical Publishing Society}, Calcutta, 478-?503.

\bibitem{} Patil, G. P., and Ord, J. K.(1975). On size-biased sampling and related form-invariant weighted distributions,  \textit{Sankhy\={a} A} Series B, 38, 48--61.

\bibitem{} Patil, G. P., and Rao, C. R.(1977). Weighted distributions: a survey of their applications. In Applications of Statistics. P. R. Krishnaiah(ed.), \textit{North Holland Publishing Company}, 383--405.

\bibitem{} Patil, G. P., and Rao, C. R.(1978). Weighted Distributions and Size-Biased Sampling with Applications to Wildlife Populations and Human Families, \textit{Biometrics}, 34(2), 179--189.

\bibitem{} Rao C.R.(1985). Weighted distributions arising out of methods of ascertainment, in A Celebration of Statistics, A.C. Atkinson \& S.E. Fienberg, eds, \textit{Springer-Verlag}, New York, Chapter 24, pp. 543-569.

\bibitem{} Sankaran M.(1969). On Certain Properties of a Class of Compound Distributions, \textit{Sankhya B}, 32, 353--362.

\bibitem{} Sankaran M.(1985). The Discrete Poisson-Lindley Distribution, \textit{Biometrika}, 26(1), 145--149.

\bibitem{} Sastry DVS, Bhati D, Rattihalli R.N. and Gómez-Déniz (2015): Zero Distorted Generalized Geometric Distribution, \textit{Communication in Statistics: Theory and Methods}. (Accepted)

\bibitem{} Tripathi, R.C., Gupta, R.C., and White, T.J. (1987). Some generalizations of the geometric distribution. Sankhy$\bar{a}$, Ser. B 49(3):218–223.

\end{thebibliography}
\end{document}